\documentclass{PoS}

\title{Inflation and Loop Quantum Cosmology}

\ShortTitle{Loop Quantum Cosmology and the Inflation}

\author{\speaker{Aurélien Barrau}\\
        Laboratoire de Physique Subatomique et de Cosmologie\\ 
	Université Joseph Fourier / CNRS / IN2P3\\ 53
	avenue des Martyrs, 38026 Grenoble cedex\\
	France\\
        E-mail: \email{barrau@in2p3.fr}}

\abstract{On the one hand, inflation is an extremely convincing scenario: it solves most
cosmological paradoxes and generates fluctuations that became
the seeds for the growth of structures. It, however, suffers from a "naturalness"
problem: generating initial conditions for inflation is far from easy. On the other
hand, loop quantum cosmology is very successful: it solves the Big Bang
singularity through a non-perturbative and background-independent quantization
of general relativity. It, however, suffers from a key drawback: it is
extremely difficult to test. Recent results can let us hope that inflation and
LQC could mutually cure those pathologies: LQC seems to naturally generate
inflation and inflation could allow us to test LQC.}

\FullConference{35th International Conference of High Energy Physics\\ ICHEP2010\\
		July 22-28, 2010\\ Paris\\ France}

\begin{document}

\section{LQC helps inflation}

Loop quantum gravity (LQG) is a tentative background-independent and non-perturbative 
quantization of general relativity. It relies on Ashtekar variables, namely
SU(2) valued connections and conjugate densitized triads. The quantization is 
performed through holonomies of the connections and fluxes of the densitized triads (see, 
{\it e.g.}, \cite{rovelli1} for excellent introductions). At the intuitive level, 
loop quantum  cosmology (LQC) can be seen as the symmetry reduced version of LQG
(although it should be underlined that the derivation of LQC from
the full theory is not yet fully demonstrated, see \cite{fran} for
the latest progresses on the spinfoam approach to LQC). While predictions of LQC
are very close to those of the Wheeler-deWitt theory in the low density
regime, there is a fundamental difference once we approach the Planck scale: 
the Big Bang is replaced by a Big Bounce due to huge repulsive quantum geometrical 
effects (see, {\it e.g.}, \cite{lqc_review} for reviews). Unquestionably, this resolution of
the primordial singularity problem is the most striking result of LQC. Following the pioneering
works \cite{bojo1}, many studies have confirmed this prediction in different situations (see, {\it e.g.},
references in \cite{ash1}).

\begin{figure}[ht!]
\centering
\includegraphics[scale=0.7]{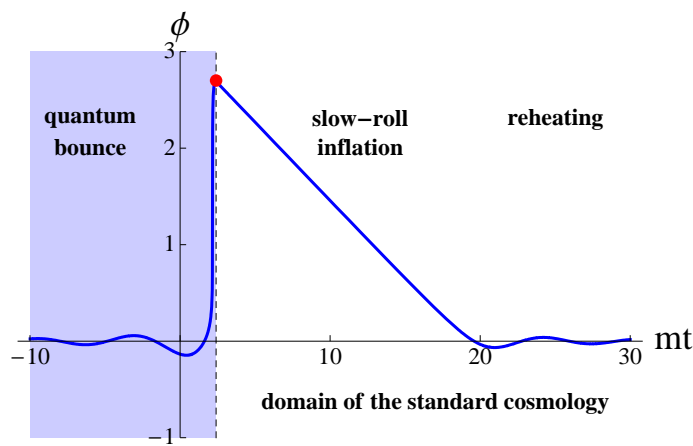} 
\caption{Evolution of a scalar field in a LQC bouncing universe. During the contraction phase
the oscillations are amplified. Then a slow-roll inflation phase takes place, followed by the reheating.}
\label{field1}
\end{figure}

Moreover, in the last years, it was realized that there are strong links between inflation and
LQC (see, {\it e.g.}, \cite{bojo2002}  for a pioneering paper on super inflation which was
followed by many others). The key feature can be understood very easily. Let's consider the
simplest possible model, without any  $\phi^4$ potential or any other intricate feature: just
a massive scale field filling the bouncing LQC universe. The Klein-Gordon equation simply reads 
$\ddot{\phi}+3H\dot{\phi}+m^2\phi=0$ where the second term is usually called the friction
term due to the expansion of the Universe. However, during the pre-bounce stage, the Universe is contracting,
the Hubble parameter is therefore negative and this becomes an anti-friction term. Otherwise
stated, the field automatically climbs-up its potential: whatever the small oscillations, they
are amplified. However, just after the bounce,
$H$ becomes positive (the Universe is expanding). This term is indeed now  a friction term and
the field is (nearly) frozen, usually high on its potential, where it was thrown away during the 
antifriction-bouncing
phase. This is {\it exactly} what is needed for slow-roll inflation to occur, without any
feature introduced "by hand" (see Fig.~\ref{field1}) ! It means
that the model naturally predicts inflation. It is rather remarkable to realize that the
canonical quantization of Einstein equations, applied to the Universe, could have predicted
inflation far before it was understood to be necessary for cosmology. 

This scenario, together with the detailed values of the parameters, is described into the
details in \cite{dreamteam1}. Furthermore, it was shown in \cite{Ashtekar:2009mm} that the
probability for a long enough inflationary phase (say with more than 60 e-folds or so)
is extremely close to 1, in sharp contrast with what was estimated for standard inflation in usual 
general relativity (\cite{turok}). In fact, as demonstrated in \cite{corichi},
it seems that this difference is mostly due to
the fact that probabilities are not estimated at the same time in the cosmic history. 
In the LQC framework, one can use a naturally defined surface and the high
probability for inflation to occur is a (quite) reliable result. My view is that
it is just not (yet) possible to define a meaningful probability for inflation 
in the standard Big Bang paradigm. This is not without echoing the problems faced for making
predictions in the multiverse...

Impressively, the bouncing scenario, as predicted in (but not only in) LQC, seems to lead
generically to inflation. Inflation is just a nearly unavoidable consequence of this model
which was, by no means, designed for this. Sounds good.

\section{Inflation helps LQC}

\begin{figure}[ht!]
\centering
\includegraphics[scale=0.7]{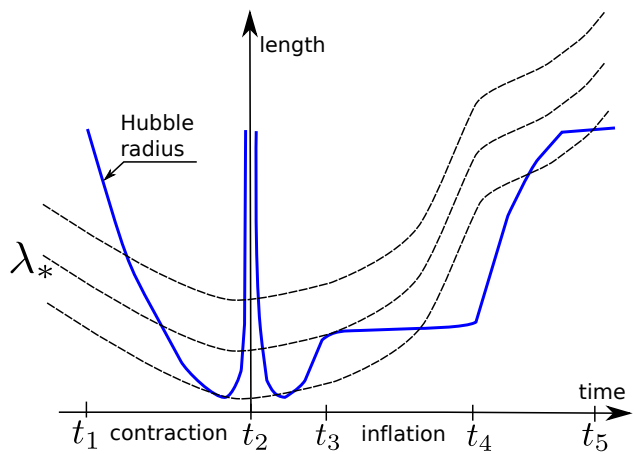}
\caption{Evolution of the Hubble radius (solid line) and of 
some length scales (dashed lines). Different times are distinguished: 
$t_1-$time when the initial conditions are set;  
$t_2-$bounce ($H=0$);  
$t_3-$beginning of inflation;  
$t_4-$end of inflation; 
$t_5-$present epoch of dark energy domination.}
\label{field2}
\end{figure}

The very exciting news is that, the other way round, inflation --which can now be considered
as quite natural-- can help us testing quantum gravity. Loop Quantum Gravity is an appealing
scenario. However, as most other attempts (including of course string theory) to quantize
gravity, it suffers from the lack of Planck-scale experiments. Measuring directly areas and
volumes at the required accuracy (to probe the discrete spectrum) is just technically 
impossible. Looking for evidences of a violation of the Lorentz invariance is an 
interesting idea. But it has so far remained inconclusive and is extremely speculative at
the theoretical level : there are no unambiguous prediction of any kind of Lorentz invariance
violation. (As pointed out by Rovelli, one should think in terms of eigenvalues of the length
operators and not in terms of usual lengths. The spectrum can remain unchanged while
the expectation value varies with speed.) Cosmology is therefore probably the best --if not only-- way to search for loopy
effects in Nature. And inflation, because it stretches  very small scales up to
macroscopic lengths, is our best friend in this game.

\begin{figure}[ht!]
\centering
\includegraphics[scale=0.6]{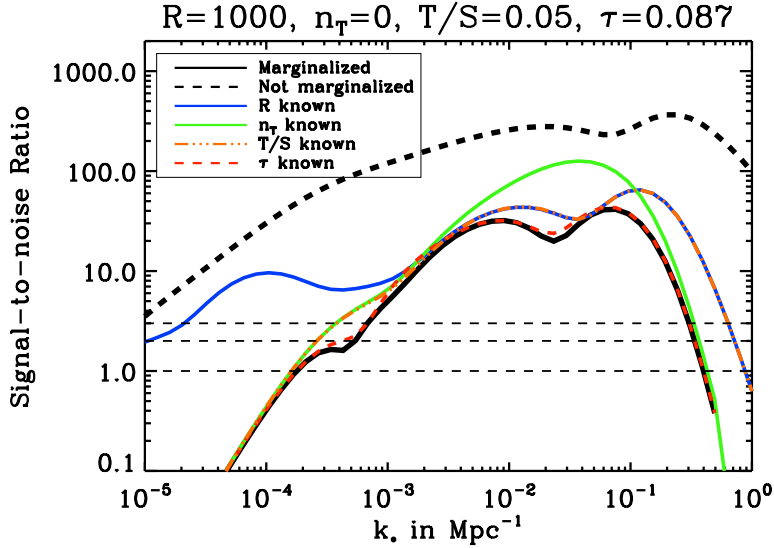}
\caption{Signal-to-noise ratio for the detection of the bounce in the CMB
B-mode for different marginalization options for a "B-POL" like mission.}
\label{field3}
\end{figure}

If one wants to be rigorous and rely on a well establish formalism, only tensor modes are
currently well understood in LQC. Primordial gravitational waves are a good candidate for
testing LQC. This has been studied in several articles (see, {\it e.g.}, \cite{lqcgen}).
The key features can be understood easily with the plot from Fig.~\ref{field2}. 
The Hubble radius ($1/H$)
is drawn in blue. The very small scales (lower black dashed curve) crosses the horizon only twice: 
they exit during inflation (the usual inflationary "plateau" reflects that $H\approx cst$) and
re-enter later on during radiation or matter domination. This is the usual picture, leading to a nearly scale-invariant power
spectrum. However, the large scales (upper black dashed curve) cross the horizon --and become frozen--
in the contracting universe, therefore exhibiting the characteristic $P(k)\propto k^2$
Minkowski vacuum spectrum. The resulting B-mode power spectrum is $k^2$ suppressed in the IR
limit and scale invariant in the UV limit. It also exhibits some oscillations between those
regimes (due to causal contact at the bounce), see~\cite{dreamteam1}.

We have recently shown (\cite{dreamteam2}) that the transition wavenumber $k_{\star}$ between 
the standard and suppressed regimes mostly depends on the {\it initial conditions} at the
bounce and {\it not} on the fundamental parameters of the theory. Basically, the wavenumber
becomes large enough to be "observable" if the Universe is strongly dominated by kinetic
energy at the bounce. This is consistent with the fact that backreaction is neglected in the
approach. It physically  means that the LQC effects can be seen if  inflation did 
not last much more than required to solve the cosmological paradoxes. There are some
arguments in favor of this, given in \cite{vega}, but they have to be taken with care. Figure~\ref{field3}
gives the result of a Fisher analysis showing the range of $k_{\star}$ values that could be
probed by the next-generation CMB experiments. The detectable range corresponds to a maximum value
of the scalar field of the order of $3.3~M_{Pl}$ for $m=10^{-6}~M_{Pl}$.

\section{Inflation and LQC live (happily?) together}

My view is that the LQC-inflation paradigm is becoming "convincing". LQC (probably) generates
inflation and inflation (possibly) allows us to test LQC. This is a tantalizing picture. Some
important points nevertheless need to be investigated. First, scalar modes (and
the resulting temperature
power spectrum of the CMB) must be studied into the details. This is on the way
(\cite{thom}) but computations are far from trivial as it is not straightforward to obtain an
anomaly-free algebra in this case.
Then, Inverse-Volume (IV) corrections should be included. All what has been said
before is related to holonomy corrections only. This should not be very difficult
and dramatic new effects are not expected as most of the observable features are
associated with the bounce itself (which will basically remain the same with IR
corrections) and not with subtle loopy corrections to the propagation of
physical modes.
Finally, and most importantly,
inhomogeneities have to be taken into account as they are known to grow very fast during the
contraction phase. This point, of course, questions the reliability of the  picture.

\end{document}